\documentclass[twocolumn]{aastex62}

\usepackage{graphicx}

\newcommand{\UDA}{Instituto de Astronom\'ia y Ciencias Planetarias, Universidad de Atacama, Copayapu 485, Copiap\'o, Chile}
\newcommand{\UTINAM}{ Institut Utinam, CNRS UMR 6213, Universit\'e Bourgogne-Franche-Comt\'e, OSU THETA Franche-Comt\'e, Observatoire de Besan\c{c}on, \\ BP 1615, 25010 Besan\c{c}on Cedex, France}
\newcommand{\UNAB}{Depto. de Cs. F\'isicas, Facultad de Ciencias Exactas, Universidad Andr\'es Bello, Av. Fern\'andez Concha 700, Las Condes, Santiago, Chile}
\newcommand{\VATICAN}{Vatican Observatory, V-00120 Vatican City State, Italy}
\newcommand{\UCN}{Instituto de Astronom\'ia, Universidad Cat\'olica del Norte, Av. Angamos 0610, Antofagasta, Chile}
\newcommand{\DAME}{Department of Physics and JINA Center for the Evolution of the Elements, University of Notre Dame, Notre Dame, IN 46556, USA}
\newcommand{\UDEC}{Departamento de Astronom\'\i a, Casilla 160-C, Universidad de Concepci\'on, Concepci\'on, Chile}
\newcommand{\SERENAa}{Department of Astronomy - Universidad de La Serena - Av. Juan Cisternas, 1200 North, La Serena, Chile}
\newcommand{\SERENAb}{Instituto de Investigaci\'on Multidisciplinario en Ciencia y Tecnolog\'ia, Universidad de La Serena. Benavente 980, La Serena, Chile}
\newcommand{\SAO}{Universidade de S\~ao Paulo, IAG, Rua do Mat\~ao 1226, Cidade Universit\'aria, S\~ao Paulo 05508-900, Brazil}    
\newcommand{\CITEVA}{Centro de Astronom\'ia (CITEVA), Universidad de Antofagasta, Av. Angamos 601, Antofagasta, Chile}
\newcommand{\UNAM}{Instituto de Astronom\'ia, Universidad Nacional Aut\'onoma de M\'exico, Apdo. Postal 106, 22800 Ensenada, B.C., M\'exico}
\newcommand{\VIRGINIA}{Department of Astronomy, University of Virginia, Charlottesville, VA 22904, USA}

\begin{document}

\title{VVV~CL001: Likely the Most Metal-Poor Surviving Globular Cluster in the Inner Galaxy}

\correspondingauthor{Jos\'e G. Fern\'andez-Trincado}
\email{jose.fernandez@uda.cl}

\author[0000-0003-3526-5052]{Jos\'e G. Fern\'andez-Trincado}
\affil{\UTINAM}
\affil{\UDA}

\author[0000-0002-7064-099X]{Dante Minniti}
\affil{\UNAB}
\affil{\VATICAN}

\author[0000-0001-8052-969X]{Stefano O. Souza}
\affil{\SAO}	

\author[0000-0003-4573-6233]{Timothy C. Beers}
\affil{\DAME}

\author{Doug Geisler}
\affil{\UDEC}
\affil{\SERENAa}
\affil{\SERENAb}

\author{Christian Moni Bidin}
\affil{\UCN}

\author{Sandro Villanova}
\affil{\UDEC}	

\author[0000-0003-2025-3147]{Steven R. Majewski}
\affil{\VIRGINIA}

\author[0000-0001-9264-4417]{Beatriz Barbuy}
\affil{\SAO}	

\author[0000-0002-5974-3998]{Angeles P\'erez-Villegas}
\affil{\UNAM}	

\author[0000-0002-2036-2944]{Lady Henao}
\affil{\UDEC}	

\author{Mar\'ia Romero-Colmenares}
\affil{\CITEVA}

\author[0000-0002-1379-4204]{Alexandre Roman-Lopes}
\affil{\SERENAa}

\author{Richard R. Lane}
\affil{\UDA}

\begin{abstract}
We present the first high-resolution abundance analysis of the globular cluster VVV~CL001, which resides in a region dominated by high interstellar reddening towards the Galactic Bulge. Using \textit{H}-band spectra acquired by the Apache Point Observatory Galactic Evolution Experiment (APOGEE), we identified two potential members of the cluster, and estimate from their Fe I lines that the cluster has an average metallicity of [Fe/H] = $-2.45$ with an uncertainty due to systematics of 0.24 dex. We find that the light-(N), $\alpha$-(O, Mg, Si), and Odd-Z (Al) elemental abundances of the stars in VVV~CL001 follow the same trend as other Galactic metal-poor globular clusters. This makes VVV~CL001 possibly the most metal-poor globular cluster identified so far within the Sun's galactocentric distance and likely one of the most metal-deficient clusters in the Galaxy after ESO280-SC06. Applying statistical isochrone fitting, we derive self-consistent age, distance, and reddening values, yielding an estimated age of $11.9^{+3.12}_{-4.05}$ Gyr at a distance of $8.22^{+1.84}_{-1.93}$ kpc, revealing that VVV~CL001 is also an old GC in the inner Galaxy. The Galactic orbit of VVV~CL001 indicates that this cluster lies on a halo-like orbit that appears to be highly eccentric. Both chemistry and dynamics support the hypothesis that VVV~CL001 could be an ancient fossil relic left behind by a massive merger event during the early evolution of the Galaxy, likely associated with either the Sequoia or the \textit{Gaia}-Enceladus-Sausage structures. 
\end{abstract}

\section{Introduction} 
\label{section1}

The inner\footnote{Here the term ``inner'' refers to objects distributed inside the Sun's galactocentric distance} Milky Way (MW) contain a significant population of globular clusters (GCs) with a wide range of metallicities ($-2.37<$ [Fe/H] $\lesssim0$; \citealt{Harris1996}, 2010 edition), most of which are still poorly explored due to the large foreground extinction and high field-star densities that complicate the analysis, especially at low Galactic latitudes. These limitations have been mitigated with wide field near-infrared imaging such as the \textit{Vista Variables in the Via Lactea} survey (VVV), and its extension the VVVX survey \citep[][]{Minniti2010, Smith2018}, which has expanded the family of Galactic GCs to more than $300$ candidates by the inclusion of objects in the inner Galaxy \citep[see e.g.][]{Minniti2017a, Minniti2017b, Palma2019, Garro2020}, including those in the bulge (Geisler et al. in prep.).

In this context, the high-resolution ($R>22,0000$) capabilities of the near-IR multi-fiber spectrographs of the Apache Point Observatory Galactic Evolution Experiment \citep[APOGEE-2;][]{Majewski2017} allow measurement of new parameters (radial velocity, metallicity, and detailed chemical abundances for many species) with high precision for a large number of Galactic GCs in a homogeneous way \citep[see e.g.][]{Meszaros2015, Schiavon2017, Masseron2019, Fernandez-Trincado2019d, Fernandez-Trincado2020d, Meszaros2020}. 

Very metal-poor GCs are preferentially associated with the oldest components of galaxies, and are often used as cosmic clocks to track the enrichment history of their host galaxy \citep{Geisler1995}. Thus, beyond the intrinsic scientific value of identifying such ancient objects, the measurements of their chemical composition can therefore provide insight into the build-up of the chemical elements in the earliest epoch of the Milky Way. 

To date, the most metal-poor MW GC known is ESO280-SC06 (located $\sim$15 kpc from the Galactic center), with an estimated metallicity of [Fe/H]$=-2.48^{+0.06}_{-0.11}$ \citep[][]{Simpson2018}, and associated to the \textit{Gaia}-Enceladus-Sausage \citep[GES;][]{Massari2019}. Here, we report another possible extreme case, VVV~CL001, which becomes possibly the most metal-poor GC known inside the Sun's Galactocentric distance, and likely the most metal-poor GC in the Galaxy. It has survived the strong tidal field of the inner MW during its interaction, and its existence implies that GCs below that of the empirical metallicity floor \citep[e.g.][]{Wan2020, Larsen2020} have been formed, but only a few exceptional cases have survived during Galactic evolution. 

In this Letter, we make use of APOGEE-2 spectra to provide the first spectroscopic study of VVV~CL001. 

\begin{figure*}[t]
	\begin{center}
		\includegraphics[height = 12. cm]{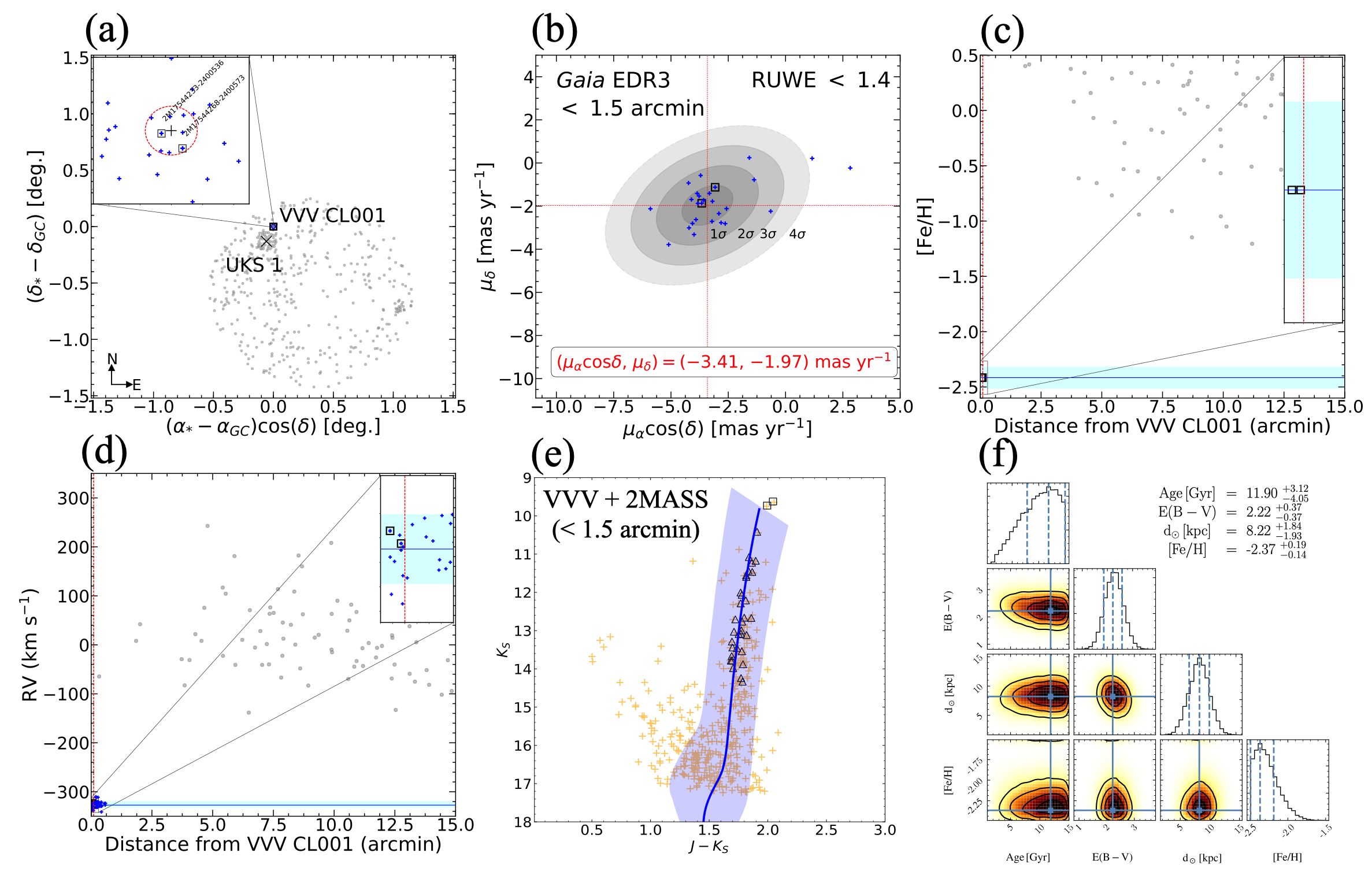}\\		
		\caption{Panel (a): The spatial distribution of APOGEE-2 stars (gray dots) observed towards VVV~CL001. The red-dashed circle highlights a radius of 0.1\arcmin\ centered on VVV~CL001 for reference. The black open squares indicate the member candidates in the innermost regions of VVV~CL001 from APOGEE-2, while the blue symbols indicate the stars with RV information compiled in \citet{Baumgardt2018}. The GC UKS~1, with its respective tidal radius, is also over-plotted for reference. Panel (b): The \textit{Gaia} EDR3 proper motion distribution of our sample. The symbols are the same as in panel (a). The concentric ellipses highlight the best-fit distribution after a 3$-\sigma$ clip, with red-dotted lines showing the best-fit proper motion of VVV~CL001, listed at the bottom of the panel. Panels (c) and (d): Metallicities and radial distribution of APOGEE-2 stars (gray dots) and Baumgardt's RV compilation (see Section \ref{section4}) in the field surrounding VVV~CL001. The blue-horizontal and shadowed-cyan region indicate the nominal metallicity and RV of VVV~CL001 within $\pm0.1$ dex and $\pm6.6$ km s$^{-1}$, respectively. The vertical red-dashed lines at 0.1\arcmin\ are highlighted for reference. Panel (e): The best isochrone fit in the $K_{s}$ versus $(J-K_s)$ CMD using DSED models, where the blue line shows the most probable solution and the blue-shadowed region indicates the solutions within 1$-\sigma$. The orange symbols mark the potential candidates from the VVV survey, while the black open squares and triangles refer to stars with RV information from APOGEE-2 and Baumgardt's compilation, respectively. Panel (f): The posterior distributions of the indicated quantities.}
		\label{Figure1}
	\end{center}
\end{figure*}

\begin{figure}[t]
	\begin{center}
		\includegraphics[height = 15. cm]{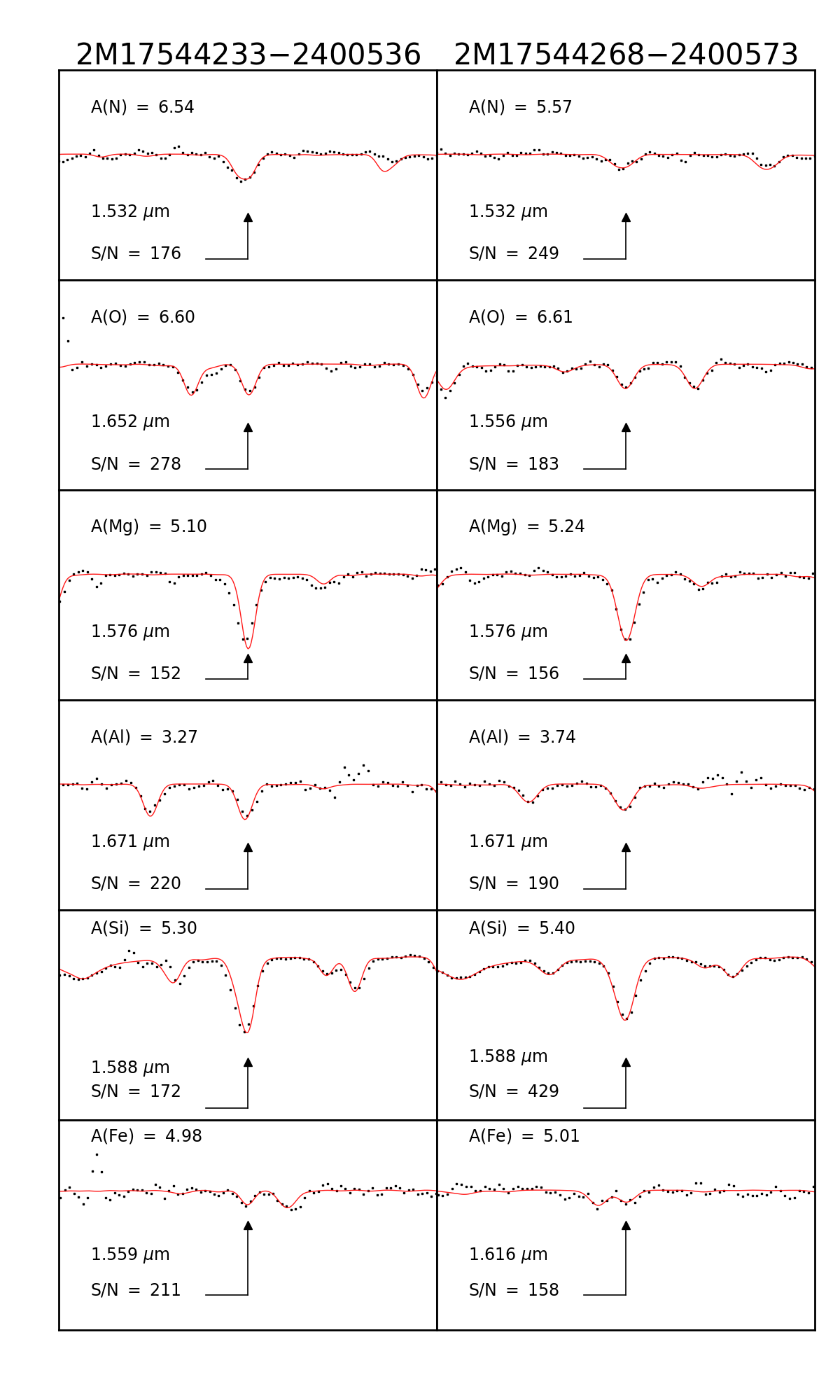}
		\caption{Detection of atomic and molecular lines. Spectral synthesis is shown for the determination of the [N/Fe], [O/Fe], [Mg/Fe], [Al/Fe], [Si/Fe], and [Fe/H] abundances for two stars in the innermost regions of VVV~CL001. Each panel shows the best-fit syntheses (red line) from \texttt{BACCHUS} compared to the observed spectra (black squares) of selected lines (black arrows).}
		\label{Figure2}
	\end{center}
\end{figure}

\begin{figure*}[t]
	\begin{center}
		\includegraphics[height = 11. cm]{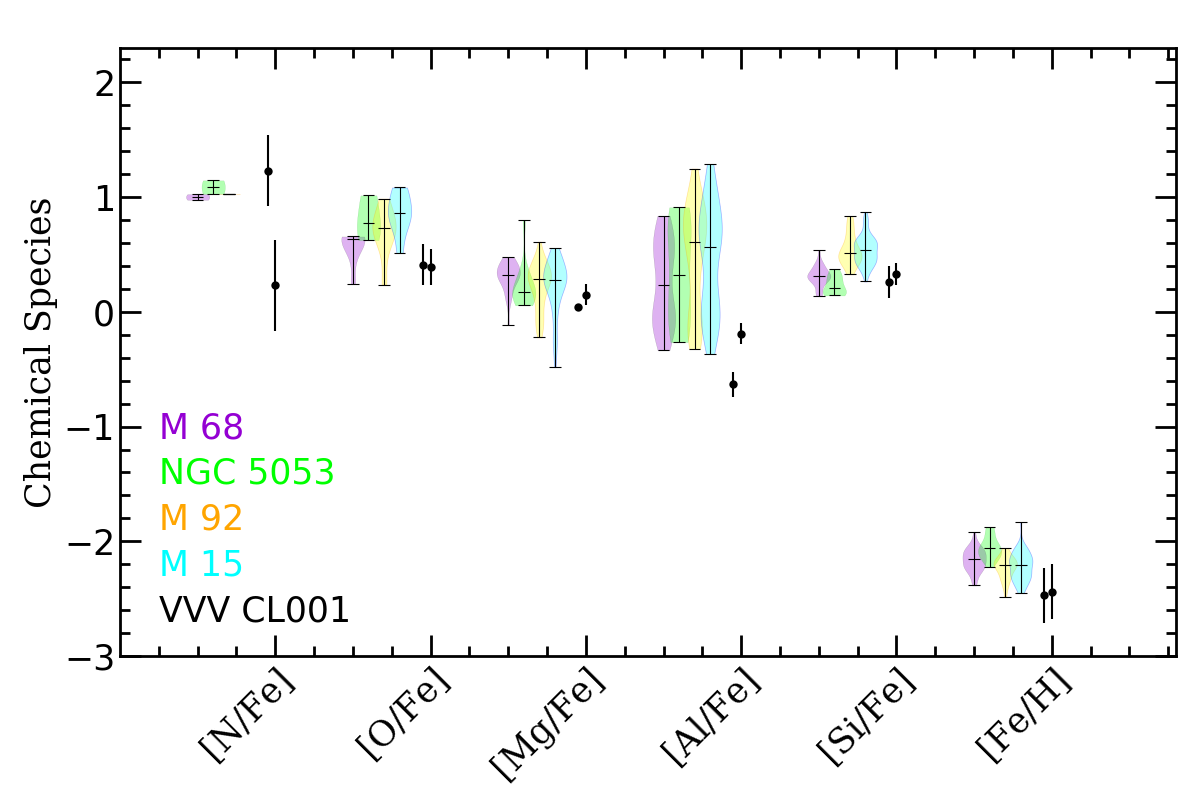}
		\caption{Elemental abundance of stars in VVV~CL001 (black symbols) and four comparison metal-poor GCs. Each violin representation show the univariate kernel density estimate of the abundance ratios of each cluster as determined from \citet{Meszaros2020}. Each GC has been slightly offset horizontally to distinguish them.}
		\label{Figure3}
	\end{center}
\end{figure*}

\begin{figure*}[t]
	\begin{center}
		\includegraphics[height = 11. cm]{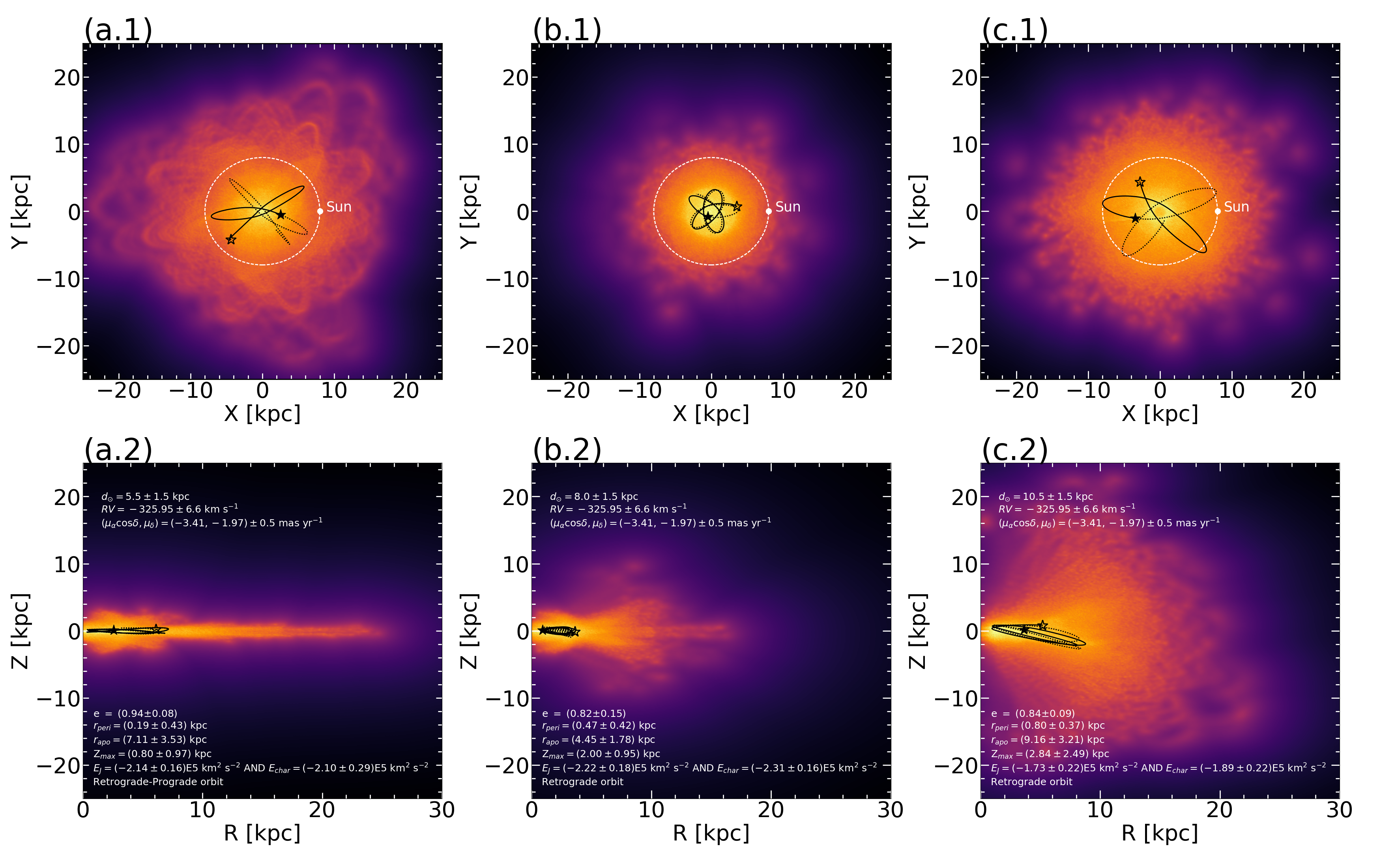}
		\includegraphics[height = 9. cm]{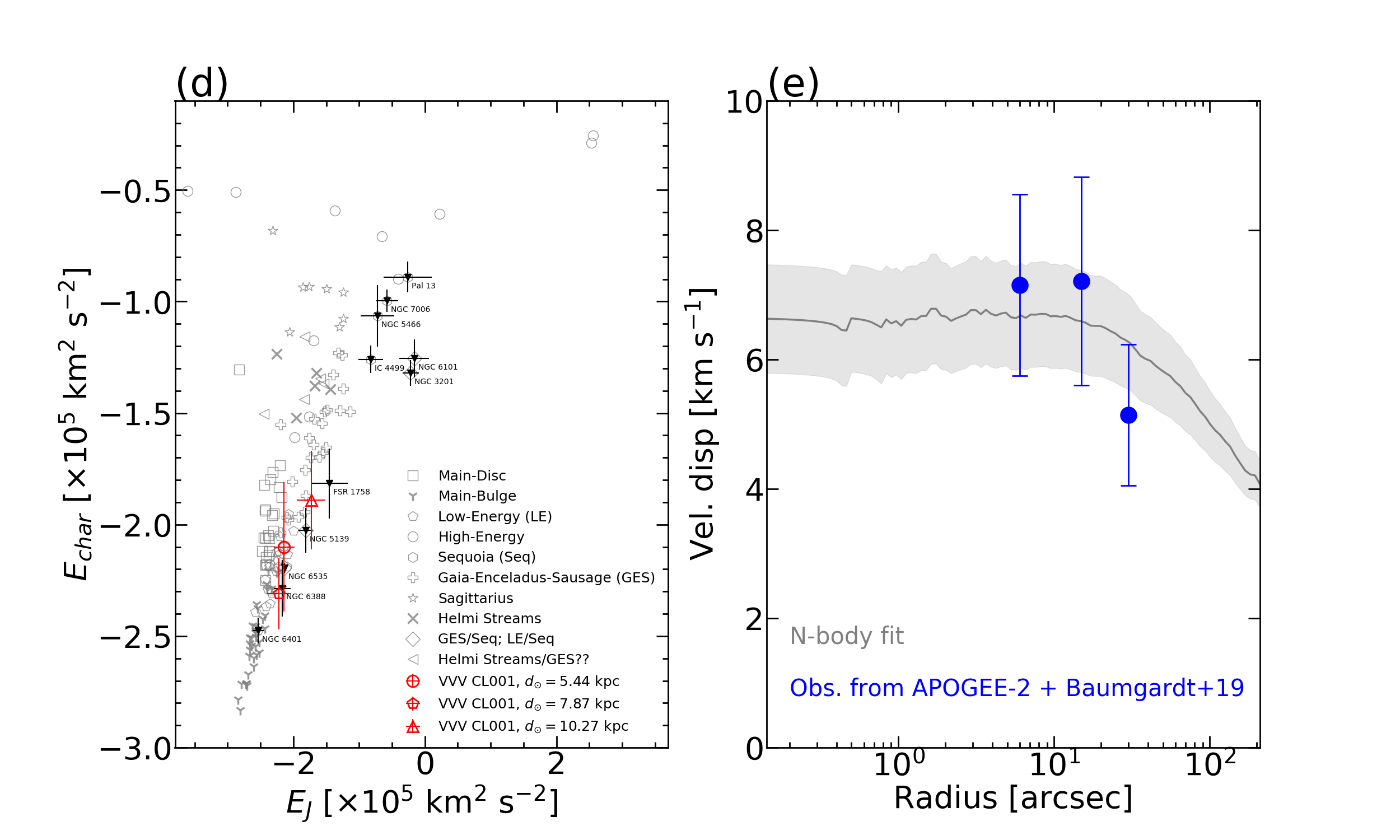}\\
		\caption{Ensemble of ten thousand orbits of VVV~CL001, projected on the equatorial (a.1, b.1, and c.1) and meridional (a.2, b.2, and c.2) Galactic planes, in the inertial reference frame with a bar pattern speed of 41 km s$^{-1}$ kpc$^{-1}$, integrated over the past 2 Gyr. The red and orange colors correspond to more probable regions of the space, which are crossed most frequently by the simulated orbits. The black solid and dashed lines show the past and future orbits of VVV~CL001, integrated over 200 Myr. The white-dashed line indicates the Sun's radius. The filled and unfilled star symbol indicate the initial and final position of the cluster, respectively. The main orbital elements are listed in panels (a.2, b.2, and c.2),  with uncertainty ranges given by the 16$^{\rm th}$ and 84$^{\rm th}$ percentile values. Panel (d): The Characteristic orbital energy ($E_{char}$) versus the orbital Jacobi constant ($E_J$) in the non-inertial reference frame where the bar is at rest. The gray symbols refer to Galactic GCs associated to different progenitors, as suggested in \citet{Massari2019}. VVV~CL001 is shown with red open symbols, while other GCs (black triangles) associated to Seq and GES are highlighted for reference. Panel (e): Line-of-sight velocity dispersion versus radius for our target cluster stars from APOGEE-2 plus Baumgardt's data set (blue dots). The prediction of the best-fitting \textit{N}-body model from \citet{Baumgardt2018} and \citet{Baumgardt2019} is shown as a solid-gray line, and the light-gray shaded region indicates the 1$-\sigma$ uncertainty from the fit.}
		\label{Figure4}
	\end{center}
\end{figure*}

\section{OBSERVATIONAL DATA} 
\label{section2}

We use an interim release data product of the APOGEE-2 survey \citep{Majewski2017} part of the Sloan Digital Sky Survey IV \citep[SDSS-IV;][]{Blanton2017} that includes data taken after the release of DR16 \citep{Ahumada2020}, to investigate, for the first time, the chemical composition of VVV~CL001. 

The (300-fiber) APOGEE instruments are high-resolution ($R\sim22,500$), near-infrared (NIR) spectrographs \citep{Wilson2019} observing all the components of the MW (halo, disc, and bulge/bar) from the Northern Hemisphere on the 2.5m telescope at Apache Point Observatory \citep[APO, APOGEE-2N;][]{Gunn2006} and from the Southern Hemisphere on the Ir\'en\'ee du Pont 2.5m telescope  \citep[][]{Bowen1973} at Las Campanas Observatory (LCO, APOGEE-2S). Each instrument records most of the \textit{H}-band (1.51$\mu$m -- 1.69$\mu$m). See \citet{Nidever2015}, \citet{Garcia2016}, and \citet{Holtzman2018} for details regarding the data reduction process and standard stellar parameter estimates. As of February 2020, the dual APOGEE-2 instruments have observed some $\sim$600,000 sources across the MW, targeting these stars according to methods described in \citet{Zasowski2013} and \citet{Zasowski2017}, with updates to the targeting plan described in Beaton et al. and Santana et al. (in prep.).

The present study focuses on VVV~CL001, a GC discovered by the VVV survey \citep[see e.g.,][]{Minniti2011}. This GC lies in the direction of the Galactic bulge, and is strongly dominated by large foreground extinction, E(B$-$V) $\gtrsim$ 2.0 mag \citep{Minniti2011}, hampering the observations of this object in the optical bands. 

VVV~CL001 stars fall on the same APOGEE-2 plug-plates as those associated with the very nearby GC UKS~1, as shown in panel (a) of Figure \ref{Figure1} \citep[see also][for details regarding to UKS~1]{Fernandez-Trincado2020d}, however the high radial velocities (RVs) of potential VVV~CL001 members allow us to cleanly distinguish the UKS~1 sources from VVV~CL001 stars. Thus, two high-confidence VVV~CL001 members, based on proper motions (see panel (b) of Figure \ref{Figure1}) from \textit{Gaia} Early Data Release 3 \citep[\textit{Gaia} EDR3:][]{Bronw2020}, location in the color-magnitude diagrams (CMDs), and RVs from APOGEE-2 were identified $\lesssim0.1'$ from the cluster centre.  

Panels (c) and (d) of Figure \ref{Figure1} show that both our derived [Fe/H] and RV of the VVV~CL001 stars are clearly distinct as compared to the foreground and background stars (hereafter field stars). The [Fe/H] and RVs are at least 1.5 dex and $\sim150$ km s$^{-1}$ offset from the field stars, respectively. Both of these offsets are very large and imply that VVV~CL001 is a truly extreme GC. The two potential cluster members from APOGEE-2 are red giant branch (RGB) stars close to the tip of the giant branch as shown in panel (e) of Figure \ref{Figure1}.

In order to have a self-consistent method for age derivation via statistical isochrone fitting, we use the \texttt{SIRIUS} code \citep[][]{Souza2020}, and the most probable cluster members in the VVV catalogue located inside 1.5$\arcmin$ from the cluster center and that have proper motions compatible with that of VVV~CL001, as well as those sources with RVs information (see Section \ref{section4}), which have been marked as triangle (for Baumgardt's data set) and square (for APOGEE-2 data set) open symbols in panel (e) of Figure \ref{Figure1}. Due to the quality of our data, we applied some assumptions to obtain an age distribution: a uniform prior in age between 1 and 15 Gyr combined with a slow drop above the age of the universe \citep[13.7 Gyr;][]{Planck2016}; the metallicity was variated around the value determined with high-resolution spectroscopy in the present work; the isochrone is limited to $\log$ g $<$ 4.5 representing the RGB region. We dereddened and extinction-corrected the VVV$+$2MASS $Ks_{s}$ and $J-K_{s}$ colors with the bulge-specific reddening maps from \citet{Gonzalez2011, Gonzalez2012} assuming the reddening law of \citet{Cardelli1989}. We also noticed that in a $\sim$2\arcmin$\times$ 2\arcmin area the differential reddening across the field do not affect the CMD of VVV~CL001, with a negligible variation of 0.03 mag in $K_{s}$. Finally, we adopted the Dartmouth Stellar Evolutionary Database \citep[DSED;][]{Dotter2008} isochrones with [$\alpha$/Fe]$=+0.4$ and canonical helium. Panel (e) and (f) of Figure \ref{Figure1} present the best isochrone fits in the K$_s$ versus ($J - K_s$) CMD. Our fit provides a reasonable solution both in the over-plotted isochrone (panel e) and the posterior distributions of the corner plot (panel f). To represent the distributions, we adopt the median as the most probable value and the uncertainties calculated from the $16^{\rm th}$ and $84^{\rm th}$ percentiles. We found an age of $11.9^{+3.12}_{-4.05}$ Gyr and a probable distance of $\sim 8.22^{+1.84}_{-1.93}$ kpc. Also, we want to stress that without the adopted assumptions, the internal error of the age determination could increase and give not a clear age distribution. Consequently, our probable solutions, within $1-\sigma$, fit well the central part of the CMD, providing confidence that the age estimate is a reasonable determination for VVV~CL001.

\section{ELEMENTAL ABUNDANCES} 
\label{section3}

As the \texttt{ASPCAP}/APOGEE-2 pipeline \citep{Garcia2016} does not provide [Fe/H] and [X/Fe] determinations for VVV~CL001 stars, we followed the same technique as described in \citet[][]{Fernandez-Trincado2019a, Fernandez-Trincado2019b, Fernandez-Trincado2019c, Fernandez-Trincado2019d, Fernandez-Trincado2020a, Fernandez-Trincado2020b, Fernandez-Trincado2020c, Fernandez-Trincado2020d}, and carried out a consistent chemical-abundance analysis for the two VVV~CL001 stars with the \texttt{BACCHUS} code \citep{Masseron2016}. The spectra of our sample, in general, have a signal-to-noise (S/N) that is appropriate for elemental-abundance determinations; see Table \ref{Table1}. The atmospheric parameters ($T_{\rm eff}$, $\log{g}$, and $\xi_{t}$), metallicity ([Fe/H]), and elemental-abundance ratios ([X/Fe]) were derived with the \texttt{BACCHUS} code. 

Figure \ref{Figure2} shows the good quality of the APOGEE-2 spectrum compared to the best-fit synthetic spectra for selected atomic and molecule lines of the two stars in VVV~CL001, from which the chemical-abundance ratios were determined. For each spectrum we were able to identify reliable lines for six chemical species (N, O, Mg, Al, Si, and Fe). The same figure also reveals the scarcely detectable Fe I lines. The resulting elemental-abundance ratios are listed in Table \ref{Table1}. 

The resulting chemical abundances are also displayed in Figure \ref{Figure3}, and compared to four GCs at similar metallicity taken from the APOGEE-2 GC sample of \citet{Meszaros2020}. We find that the two program stars exhibit an iron abundance ratio of $-2.47$ to $-2.44$, suggesting that VVV~CL001 has a mean metallicity [Fe/H] $= -2.45$, with an uncertainty due to systematics of 0.24 dex, which makes this cluster possibly the most metal-poor GC identified so far within the Sun's Galactocentric distance, and with an extreme metallicity close to the apparent ``floor" in the empirical metallicity distribution function for GCs in the MW  and Local Universe \citep[e.g.,][]{Geisler1995, Simpson2018, Kruijssen2019, Larsen2020, Wan2020}. Note that VVV~CL001 has the lowest Fe abundance amongst this sample, which includes the lowest metallicity GCs observed by APOGEE. With our limited sample, we do not find evidence for an intrinsic Fe-abundance spread. 

Regarding the $\alpha$-elements (O, Mg, and Si ), VVV~CL001 displays a modest $\alpha$-element enhancement, a clear signature of the fast enrichment provided by supernovae (SNe) II events, and compatible with other Galactic metal-poor GCs at similar metallicity (see Figure \ref{Figure3}). Furthermore, no Mg-Al anti-correlation is evident in our small sample; the two VVV~CL001 stars exhibit an aluminum deficit, which places them within the definition of \textit{first-generation} stars according to the criteria developed by \citet{Meszaros2020}. However, one star in our sample displays a very high enrichment in nitrogen, and is considered to likely belong to some of the families of \textit{second-generation}\footnote{\textit{Second-generation} is used here to refer to stars in VVV~CL001 that display altered light-element abundances (e.g. He, C, N, O, Na, Al, and Mg), which are different from those of typical MW field stars.} stars with low-aluminum enrichment \citep[see e.g.][]{Meszaros2020}, indicating the possible evidence for multiple stellar populations (MPs) in VVV~CL001. However, we caution the reader that the error bars in either N or Al abundance (or both) are not properly understood or inconclusive, therefore more cluster members need to be followed-up of in order to confirm or discard the existence of MPs in this VVV~CL001.

\begin{table}
		\begin{center}
			\setlength{\tabcolsep}{3.0mm}  
			\caption{Atmospheric Parameters, and Elemental Abundances of Stars in VVV~CL001}
			\begin{tabular}{|l|c|c|}
				\hline
				APOGEE$-$ID       &  2M17544233  & 2M17544268\\ 
				      & $-$2400536   & $-$2400573 \\ 
				\hline
				$T_{\rm eff}$ (K)      &      4100                          &           4200                      \\
				$\log$ \textit{g} (cgs) &      0.50                           &           0.50                       \\
				$\xi_{t}$ (km s$^{-1}$)            &      1.81                            &          1.69                          \\
				S/N (pixel$^{-1}$)                &      229                             &        193                                 \\
				 RV (km s$^{-1}$)               &      $-323.57$                   &    $-326.16$                           \\
				 RV-scatter (km s$^{-1}$)               &      $2.55$                   &    $2.13$     \\				 
				${\rm [N/Fe]}$                  &  $+1.23 \pm 0.31$         & $+0.23 \pm 0.40$       \\
				${\rm   [O/Fe]}  $                &  $+0.41 \pm 0.18$        &  $+0.39 \pm 0.16$       \\
				 ${\rm  [Mg/Fe]} $               &  $+0.04 \pm 0.02$        &  $+0.15 \pm 0.09$       \\
				  ${\rm  [Al/Fe]} $               &   $-0.63 \pm 0.11$        &  $-0.19 \pm 0.09$       \\
				  ${\rm  [Si/Fe]} $               &    $+0.26 \pm 0.14$        & $+0.33 \pm 0.10$       \\
				  ${\rm  [Fe/H]}$                &    $-2.47 \pm 0.24$         & $-2.44 \pm 0.24$       \\
				\hline
			\end{tabular}  \label{Table1}
				\tablecomments{ 
							The errors were determined in the same manner, as described in \citet{Fernandez-Trincado2020a} by variating the atmospheric parameters one at a time by the typical, albeit conservative, values of $\Delta T_{\rm eff} = \pm 100$ K, $\Delta \log$ \textit{g} $= \pm 100$ cgs, and $\Delta \xi_t = \pm 0.05$ km s$^{-1}$. Thus, the reported uncertainties are defined as $\sigma^{2}_{total}  = \sigma^2_{[X/H], T_{\rm eff}}    + \sigma^2_{[X/H],{\rm log} g} + \sigma^2_{[X/H],\xi_t}  + \sigma^2_{mean}$.
		}	
		\end{center}
\end{table}     

\section{DYNAMICAL PROPERTIES} 
\label{section4}

We made use of the state-of-art MW model  --\texttt{GravPot16} to predict the orbital path of VVV~CL001 in a steady-state gravitational Galactic model that includes a ``boxy/peanut" bar structure. 

For the orbit computations, we adopt the same Galactic model configuration, solar position and velocity vector as described in \citet{Fernandez-Trincado2020e}, except for the angular velocity of the bar, for which we employed the recommended value of 41 km s$^{-1}$ kpc$^{-1}$ \citep[][]{Sanders2019}. 

The most likely orbital parameters and their uncertainties are estimated using a simple Monte Carlo approach.  An ensemble of ten thousand orbits were calculated under variations of the observational parameters according to their estimated errors (assumed as 1$-\sigma$ variation), where the errors were assumed to follow a Gaussian distribution. 

To compute the orbits, we adopt a mean RV of $-325.95 \pm 6.6$ km s$^{-1}$ computed from the combined APOGEE-2 RVs of the two program stars and Holger Baumgardt's RV compilation\footnote{\url{https://people.smp.uq.edu.au/HolgerBaumgardt}} of 34 stars, which lie in the  RV range between $-341.13$ to $-310.22$ km s$^{-1}$. From those stars, we select objects having a Renormalized Unit Weight Error (\texttt{RUWE}) below 1.4, as extracted from \textit{Gaia} EDR3, which allows us to discard sources with problematic astrometric solutions \citep[see e.g.][]{Lindegren2018}. This reduces the 36 stars with RV information to 27 stars with both reliable proper-motions and RV information, which were considered to compute the nominal proper motion of VVV~CL001 after applying a 3$-\sigma$ clipping to the data, e.g., only 25 out of these 27 sources lies inside 3$-\sigma$ of the nominal proper motion of the cluster, as shown in panel (b) of Figure \ref{Figure1} . From this procedure, the nominal proper motion of VVV~CL001 is $(\mu_{\alpha}\cos(\delta), \mu_{\delta})= (-3.41, -1.97)$ mas yr$^{-1}$, with an assumed uncertainty of 0.5 mas yr$^{-1}$. The heliocentric distance ($d_{\odot}$) of VVV~CL001 remains uncertain; for this reason we assume three possible estimate (5.5, 8.0, and 10.5 kpc) close to the best isochrone fit, as shown in panel (f) of Figure \ref{Figure1}.

The main orbital elements and the ensemble of orbits of VVV~CL001 are displayed in panels (a) to (c) and (d) of Figure \ref{Figure4}. The orbital parameters reveal that VVV~CL001 lies on a radial and highly eccentric ($>$0.8) halo-like orbit with rather small excursions above the Galactic plane ($Z_{max} <$ 3 kpc), and pericentric ($r_{peri}$) distances below 1 kpc. We also find that VVV~CL001 exhibits a retrograde and prograde sense at the same time when assuming a close heliocentric distance, while it is exclusively retrograde at and beyond the Bulge, similar to other GC GCs in the inner Galaxy \citep[see][]{Perez-Villegas2020}. Therefore, a more robust heliocentric distance estimation will better constraint these dynamics scenarios.   

Panel (d) of Figure \ref{Figure4} shows that the orbital energy configuration of VVV~CL001 is comparable to that of Galactic GCs associated with the major accretion events such as Sequoia (Seq.) and GES \citep[see e.g.,][]{Myeong2018, Massari2019}, indicating that VVV~CL001 could be the fossil relic of one of these accreted dwarf galaxies. 

\section{Mass} 
\label{section5}

\citet{Baumgardt2018} performed \textit{N}-body simulations of star clusters, and found that they could reproduce the surface-density profile of VVV~CL001, finding a present-day mass of 9$\times$10$^{4}$ M$_{\odot}$. 

With the available RV data, we match the line-of-sight dispersion profiles to the \textit{N}-body simulations of  VVV~CL001, as shown in panel (e) of Figure \ref{Figure4}, and thus determine the most likely mass of the cluster from kinematics constraint. We adopted three radial bins (with bin centers of 6\arcmin, 15\arcmin, and 30\arcmin), and chosen to ensure that at least ten stars were in each bin; resulting in the three blue dots shown in panel (e) of \ref{Figure4}. With the new data in panel (e) of Figure \ref{Figure4}, we find $\sigma_{0}\sim6.6$ km s$^{-1}$. This yields a present-day estimated mass of $\sim$2.1$\times$10$^{5}$ M$_{\odot}$, which suggests that VVV~CL001 is two times more massive than previously thought, and approximately four times more massive than ESO280-SC06 (the most metal-poor GC known in the MW).

\section{CONCLUDING REMARKS} 
\label{section6}
 
 We have performed the first near-IR high-resolution spectral analysis of two likely members of the GC VVV~CL001, a cluster obscured by the heavy extinction and high field-star density in the direction of the Galactic Bulge. Based on high S/N APOGEE spectra, we measure a mean [Fe/H] metallicity of $-2.45\pm0.1$, which makes VVV~CL001 possibly the most metal-poor GC known, only slightly higher than ESO280-SC06 at $-2.48$.
 
VVV~CL001 is very close in projection to UKS~1, which motivated \citet{Minniti2011} to suggest the possibility that they could be gravitationally bound. However, the RV of VVV~CL001 is too large, and the orbits are very different, which allows us to rule out the binary-cluster scenario.

We find that the $\alpha$-element abundances of VVV~CL001 are typical of the lowest metallicity GCs known. Spectra for more members are required in order to confirm the presence of MPs in this cluster.

For the derivation of the age and reddening, we employed the new code \texttt{SIRIUS} \citep{Souza2020}. As shown in panel (e) of Figure \ref{Figure1}, we derived a median age of $\sim11.9^{+3.12}_{-4.05}$ Gyr, indicating that the very metal-poor GC VVV~CL001 is among the oldest and most massive ($\sim$2.1$\times$10$^{5}$ M$_{\odot}$) MW clusters.
   
A dynamical analysis of VVV~CL001 reveals that this object has a radial and highly eccentric halo-like orbit confined inside the Sun's galactocentric distance. Both its metallicity and orbit favor the interpretation of VVV~CL001 being a GC that belongs to an early accretion event in the MW corresponding to the either the Seq or GES dwarf galaxies. Finally, multi-band photometry in the near-IR will be useful to identify other stellar tracers, such as RR Lyrae stars (if any) toward VVV~CL001, which will significantly help to constrain the distance, age, and origin of the cluster.
  
\acknowledgments
We thank the anonymous referee for helpful comments that greatly improved the paper. We warmly thank Holger Baumgardt for providing his published numerical \textit{N}-body modeling of the line-of-sight velocity dispersion of VVV~CL001.  
J.G.F-T is supported by FONDECYT No. 3180210. 
D.M. is supported by the BASAL Center for Astrophysics and Associated Technologies (CATA) through grant AFB 170002, and by project FONDECYT Regular No. 1170121. 
S.O.S. acknowledges the FAPESP PhD fellowship 2018/22044-3. 
T.C.B. acknowledges partial support for this work from grant PHY 14-30152: Physics Frontier Center / JINA Center for the Evolution of the Elements (JINA-CEE), awarded by the US National Science Foundation. 
D.G. gratefully acknowledges support from the Chilean Centro de Excelencia en Astrof\'isica y Tecnolog\'ias Afines (CATA) BASAL grant AFB-170002. D.G. also acknowledges financial support from the Direcci\'on de Investigaci\'on y Desarrollo de la Universidad de La Serena through the Programa de Incentivo a la Investigaci\'on de Acad\'emicos (PIA-DIDULS). 
S.V. gratefully acknowledges the support provided by Fondecyt reg. n. 1170518. 
B.B. acknowledge partial financial support from FAPESP, CNPq, and CAPES - Finance Code 001. 
A.P-V. and S.O.S acknowledge the DGAPA-PAPIIT grant IG100319.
L.H. gratefully acknowledges support provided by National Agency for Research and Development (ANID)/CONICYT-PFCHA/DOCTORADO NACIONAL/2017-21171231. 
A.R-L. acknowledges financial support provided in Chile by Comisi\'on Nacional de Investigaci\'on Cient\'ifica y Tecnol\'ogica (CONICYT) through the FONDECYT project 1170476 and by the QUIMAL project 130001. 
\newline
Funding for the Sloan Digital Sky Survey IV has been provided by the Alfred P. Sloan Foundation, the U.S. Department of Energy Office of Science, and the Participating Institutions. SDSS-IV acknowledges support and resources from the Center for High-Performance Computing at the University of Utah. The SDSS website is www.sdss.org.
\newline
SDSS-IV is managed by the Astrophysical Research Consortium for the Participating Institutions of the SDSS Collaboration including the Brazilian Participation Group, the Carnegie Institution for Science, Carnegie Mellon University, the Chilean Participation Group, the French Participation Group, Harvard-Smithsonian Center for Astrophysics, Instituto de Astrof\'{i}sica de Canarias, The Johns Hopkins University, Kavli Institute for the Physics and Mathematics of the Universe (IPMU) / University of Tokyo, Lawrence Berkeley National Laboratory, Leibniz Institut f\"{u}r Astrophysik Potsdam (AIP), Max-Planck-Institut f\"{u}r Astronomie (MPIA Heidelberg), Max-Planck-Institut f\"{u}r Astrophysik (MPA Garching), Max-Planck-Institut f\"{u}r Extraterrestrische Physik (MPE), National Astronomical Observatory of China, New Mexico State University, New York University, the University of Notre Dame, Observat\'{o}rio Nacional / MCTI, The Ohio State University, Pennsylvania State University, Shanghai Astronomical Observatory, United Kingdom Participation Group, Universidad Nacional Aut\'{o}noma de M\'{e}xico, University of Arizona, University of Colorado Boulder, University of Oxford, University of Portsmouth, University of Utah, University of Virginia, University of Washington, University of Wisconsin, Vanderbilt University, and Yale University.
\newline
This work has made use of data from the European Space Agency (ESA) mission \textit{Gaia} (\url{http://www.cosmos.esa.int/gaia}), processed by the \textit{Gaia} Data Processing and Analysis Consortium (DPAC, \url{http://www.cosmos.esa.int/web/gaia/dpac/consortium}). Funding for the DPAC has been provided by national institutions, in particular the institutions participating in the \textit{Gaia} Multilateral Agreement.	
\newline
Simulations have been executed on HPC resources on the Cluster Supercomputer Atocatl from Universidad Nacional Aut\'onoma de M\'exico (UNAM). 


\end{document}